# Quantum phase transition of correlated iron-based superconductivity in LiFe$_{1-x}$Co$_x$As

**Authors:** Jia-Xin Yin[1]*†, Songtian S. Zhang[1]*, Guangyang Dai[2]*, Yuanyuan Zhao[3]*, Andreas Kreisel[4]*, Gennevieve Macam[5]*, Xianxin Wu[2], Hu Miao[6], Zhi-Quan Huang[5], Johannes H. J. Martiny[7], Brian M. Andersen[8], Nana Shumiya[1], Daniel Multer[1], Maksim Litskevich[1], Zijia Cheng[1], Xian Yang[1], Tyler A. Cochran[1], Guoqing Chang[1], Ilya Belopolski[1], Lingyi Xing[2], Xiancheng Wang[2], Yi Gao[9], Feng-Chuan Chuang[5], Hsin Lin[10], Ziqiang Wang[11], Changqing Jin[2], Yunkyu Bang[12], M. Zahid Hasan[1,13]†

**Affiliations:**
[1]Laboratory for Topological Quantum Matter and Advanced Spectroscopy (B7), Department of Physics, Princeton University, Princeton, New Jersey, USA.

[2]Institute of Physics, Chinese Academy of Sciences, Beijing, China.

[3]School of Physics and Optoelectronic Engineering, Nanjing University of Information Science and Technology, Nanjing, China.

[4]Institut für Theoretische Physik, Universität Leipzig, Leipzig, Germany.

[5]Department of Physics, National Sun Yat-Sen University, Kaohsiung, Taiwan.

[6]Brookhaven National Laboratory, Upton, New York, USA.

[7]Center for Nanostructured Graphene (CNG), Dept. of Micro- and Nanotechnology, Technical University of Denmark, DK-2800 Kongens Lyngby, Denmark.

[8]Niels Bohr Institute, University of Copenhagen, Copenhagen, Denmark.

[9]Center for Quantum Transport and Thermal Energy Science, Jiangsu Key Lab on Opto-Electronic Technology, School of Physics and Technology, Nanjing Normal University, Nanjing, China.

[10]Institute of Physics, Academia Sinica, Taipei, Taiwan.

[11]Department of Physics, Boston College, Chestnut Hill, Massachusetts, USA.

[12]Asia Pacific Center for Theoretical Physics and Department of Physics, POSTECH, Pohang, Gyeongbuk, Korea.

[13]Lawrence Berkeley National Laboratory, Berkeley, California, USA.

†Corresponding author, E-mail: mzhasan@princeton.edu; jiaxiny@princeton.edu

*These authors contributed equally to this work.

**Abstract:** The interplay between unconventional Cooper pairing and quantum states associated with atomic scale defects is a frontier of research with many open questions. So far, only a few of the high-temperature superconductors allow this intricate physics to be studied in a widely tunable way. We use scanning tunneling microscopy (STM) to image the electronic impact of Co atoms on the ground state of the $LiFe_{1-x}Co_xAs$ system. We observe that impurities progressively suppress the global superconducting gap and introduce low energy states near the gap edge, with the superconductivity remaining in the strong-coupling limit. Unexpectedly, the fully opened gap evolves into a nodal state before the Cooper pair coherence is fully destroyed. Our systematic theoretical analysis shows that these new observations can be quantitatively understood by the nonmagnetic Born-limit scattering effect in a $s\pm$-wave superconductor, unveiling the driving force of the superconductor to metal quantum phase transition.

In the research of high-$T_c$ superconductors, chemical substitution is a powerful way to manipulate electronic phases [1-5]. Meanwhile, chemical substitution also creates imperfections at the atomic scale, which break the unconventional Cooper pairing [4,5]. Although the single atomic impurity pair-breaking effect has been demonstrated in certain superconducting systems [4,5], it is challenging to study its collective many-body manifestation (the finite-density-impurity problem) in a widely tunable way, due to the existence of competing orders or inhomogeneity from strong electron correlation [1-5]. In this regard, the $LiFe_{1-x}Co_xAs$ is a rare case in which Co substitution monotonically suppresses the homogeneous superconductivity in LiFeAs without generating other competing orders [6-12], making it a versatile platform to quantitatively test many-body theories. Intriguingly, photoemission, optical and magnetic response experiments [7-11] reveal that Co substitution changes the Fermi surface and enhances the Fermi surface nesting along with the associated low-energy spin fluctuation, while the spin fluctuation is generally believed to be beneficial for the Cooper pairing in this material [13-15]. This contrast implies a striking, yet not understood de-pairing mechanism associated with Co substitution. Unexpectedly, previous STM experiments found no detectable local pair-breaking effects associated with a single Co impurity [16,17]. There is also no direct spectroscopic data measured deep in the superconducting state demonstrating how a finite density of Co impurities collectively suppresses Cooper pairing. Therefore, a systematic microscopic examination of the effect of the Co substitution on the ground state of $LiFe_{1-x}Co_xAs$ across the whole superconducting phase diagram is demanded.

LiFeAs crystallizes in a tetragonal unit cell (P4/nmm) as shown in Fig. 1(a) with a superconducting transition temperature $T_C$ of ~17K. In momentum space, it features hole-like Fermi surfaces at the Brillouin zone center and electron-like Fermi surfaces around the zone boundary, with two extra Dirac cones at the zone center being recently observed [12] (Figs. 1(b)). We first probe the superconducting ground state of the pristine material at T = 0.4K. Our atomically resolved high resolution STM image reveals a tetragonal lattice which is the Li-terminating surface (Fig. 1(c)). A line-cut of the differential conductance spectra probing the local density of state (DOS) shows a spatially homogenous double-gap structure, with a larger gap of 6.0meV and a smaller gap of 3.3meV (Fig. 1(d)). Based on previous photoemission data [18] measured at 8K, the large gap likely arises from the electron bands and the inner hole-like band, and the smaller gap likely arises from the outer hole-like band.

As the Fe lattice is systematically substituted with Co atoms, the $T_C$ decreases linearly and reaches zero around x = 16% (Fig. 2(a)) [6-11]. Based on the photoemission data [12], the Fermi level can be systematically tuned by increasing Co concentration as illustrated in the inset of Fig. 2(a). Upon bulk substitution of 1% Co atoms, STM topographical scans reveal new dumbbell-like defects randomly scattered on the surface (Fig. 2(b)) that are different from various native defects in LiFeAs. The concentration of these defects is consistent with the nominal Co substitution. The dumbbell-like defects are also randomly aligned along two orthogonal directions, with its local two-fold symmetry arising from the structural geometry. The center of each such defect is located at the middle of two Li atoms (Fig. 2(c)), which corresponds to the position of the Co atom in the underlying (Fe, Co) lattice (Fig. 2(d) inset), and altogether they possess a local two-fold symmetry. Thus, these defects are likely caused by the atomic Co substitution [17]. Directly above these dumbbell defects, we observe a state near the smaller gap at the positive energy while the overall gap structure remains almost unchanged compared with the far away spectrum (Fig. 2(d)). The weak in-gap state is consistent with earlier calculations [19] based on the band structure and impurity potentials of Co obtained from density functional theory. We note that the observation of the small local electronic variation may benefit from our lower temperature (0.4K) and more dilute impurity concentration compared with previous STM studies [16,17]. Our observation indicates that the dilute Co substitution has a limited local impact on the superconducting order parameter or causes only very weak pair-breaking scattering.

With increasing Co concentration, the Co induced weak in-gap states overlap spatially, making them difficult to be visualized individually [16,17]. On the other hand, the finite concentration of Co impurities collectively suppresses bulk superconductivity. To study the global effects on the superconducting ground state, we systematically probe the spectra away from the apparent surface defects for a wide range of Co concentrations at base temperature 0.4K. We observe a strong variation of the superconducting gap structure in the tunneling conductance which correlates strongly with the $T_C$ reduction (Fig. 2(e)). As the Co concentration increases, the large superconducting gap size decreases progressively until no gap remains at x = 16% where $T_C$ = 0. Meanwhile, the superconducting coherence peak grows progressively weaker. Evidently, the spectral bottom evolves from a U-shape to a V-shape and then gradually elevates to the normal state value.

The Co induced gap reduction and scattering can also be qualitatively reflected in the vortex excitation. We extensively study the vortices (Fig. 3) for different Co concentrations at 0.4K with c-axis magnetic fields. In the pristine sample (Fig. 3 (a)), the vortices form an ordered hexagonal lattice under a zero-field cooling method [20,21], as can be clearly seen in the autocorrelation of the real-space mapping at 2T (Fig. 3(a) inset). As the Co concentration x increases, we find the vortex lattice symmetry to remain hexagonal like (Figs. 3(b) inset), while the vortex core size increases. The persistent hexagonal vortex lattice symmetry indicates that the randomly distributed Co dopants do not distort the vortex lattice significantly. As the core size is related with the coherence length which is proportional to the reverse of the gap in the BCS theory, the increment of the vortex core size is consistent with the aforementioned gap reduction. Moreover, measuring the conductance within a vortex under an applied c-axis field of 0.5T reveals sharp in-gap bound states at $|E| \approx 1$meV (Figs. 3(b)) [20,21], in agreement with the estimate of vortex core states energies in the quantum limit, which should be on the order of a non-topological superconducting vortex state (in the energy order of $\pm\Delta^2/E_F$). As the doping concentration increases, these sharp bound states become gradually less pronounced (Figs. 3(b)), consistent with the aforementioned increased scattering. For each concentration, we carefully examine at least six vortex core states, but do not find any that exhibits a pronounced zero-energy peak. This absence of localized zero-energy states is consistent with the detailed band topology of $LiFe_{1-x}Co_xAs$. According to the photoemission study [12] and first-principles calculations (Fig. 1(b) and Fig. 2(a) inset), the surface Dirac cone (lower cone) is buried below the Fermi level in the three-dimensional bulk states, and hence does not form surface helical Cooper pairing and distinct Majorana bound states localized at the ends of the vortex line [22]. Moreover, the expected spectra of the vortex lines in superconductors with bulk Dirac states are not yet fully understood. Recently, there have been theoretical studies of the expected Majorana modes [23,24]. However, details of the vortex properties leave the possibility that these states are not localized at the vortex ends and the system might not feature zero energy bound states. These conclusions are not inconsistent with our experimental data, and we want to stress that it is a challenge to unambiguously distinguish the non-localized Majorana state by STM technique alone [23,24].

To quantify the Co induced gap reduction and scattering, we extract two key parameters from the raw data: the large energy gap size $\Delta_L$ and global zero-energy density of state $N(E=0)$. Remarkably, we find that $\Delta_L$ decreases linearly as a function of x and reaches zero around 16%, which scales linearly with the reducing $T_C$ (Fig. 4(a)). In other words, the coupling strength $2\Delta_L/k_BT_C$ remains a constant (inset of Fig. 4(a)). In particular, $LiFe_{1-x}Co_xAs$ remains in the strong coupling limit for all x as evidenced by $2\Delta_L/k_BT_C \approx 7.7$, much larger than the BCS value 3.5. These results suggest that the superconductivity is destroyed via a mechanism which decreases the pair susceptibility strength, but not the coupling strength. On the other hand, the extracted zero-energy state $N(E=0)$ exhibits an exponential like growth as shown in Fig. 4(b). The comparatively smaller rate of growth increase of $N(E=0)$ at low concentrations is consistent with the local effect of each Co atom individually (Fig. 2(d)) that each Co induces weak impurity state near the superconducting gap edge (Fig. 2(d)). As the concentration increases, the interference of their impurity wave functions becomes stronger and the global impurity states spread further in energy, and their tail states eventually contribute to the rapid rise of the global zero-energy state.

In our systematic first-principles calculations, we find that the Co dopants are essentially nonmagnetic with a relatively weak on-site potential of -0.43eV (Supplementary), consistent with previous experiments showing that they do not introduce a local magnetic moment [6,10,11,25]. According to the Anderson theorem, nonmagnetic impurities have little effect on the conventional s-wave superconductor. With a sign change in the order parameter, nonmagnetic impurity is then able to break the Cooper pairs [4,5,26-29]. Considering previous phase sensitive experiments [21] in this compound, the strongest pairing wave-function candidate is s± (where the sign changes between the ordinary hole and electron Fermi surfaces). Crucially, the variation of the gap structure from U-shape to V-shape due to nonmagnetic scattering in the s± pairing state has been predicted using the T-matrix theory [26]. Taking this two-band model from Ref. 26, we set both linear gap reduction and linear scattering rate enhancement with increasing x (Supplementary), and compute $N(E=0)$ under the Born (weak scattering) limit and the unitary (strong scattering) limit [4,5] with the results shown in Fig. 4(b). We find that the experimental data is consistent with the former condition and deviates substantially from the latter. Figure 4(d) displays the calculated DOS in the Born limit, which gradually evolves from a fully opened gap to a less coherent V-shaped structure, in consistency with our experimental observation (Fig. 2(e)). In this model, such behavior is due to the impurity states residing near the gap edge (which can be qualitatively identified from the imaginary part of the quantum many-body self-energy, as detailed in supplementary) with their tail states gradually moving towards zero-energy. Therefore, this theory offers a heuristic understanding of our experiment, demonstrating the Born limit nonmagnetic scattering nature of Co and sign reversal of the gap symmetry.

To acquire a self-consistent and quantitative understanding of the quantum many-body effect of the Co dopants, we further perform real-space calculations using the Bogoliubov–de Gennes (BdG) approach. We first take a two-orbital effective model capturing the essence of its low energy multi-band structure and consider randomly distributed electron dopants with weak potential scattering as Co impurities in reference to first-principles calculation (Supplementary). The next-nearest-neighbor intra-orbital attraction is considered to cause the s± wave Cooper pairing. The calculated DOS indeed shows a clear U-shape to a V-shape evolution as demonstrated in Fig. 4(f). This encourages us to further perform a fully realistic calculation with complete five-orbitals. The five-orbital model has successfully explained the vortex core states[20] and weak Co impurity states in pristine LiFeAs[16,19], where the s± wave Cooper pairing is self-consistently obtained within spin-fluctuation mediated pairing. Considering similarly weak potential scattering, the calculated DOS and phase diagram are shown in Fig. 4(e), which reasonably agrees with the experiment in realistic energy units. We stress that the latter five-band theoretical study contains no free fitting parameters since the band, gap structure, and impurity potential are fixed by either experiment or first-principles calculations. In this respect, it constitutes a new level of quantitative disorder modelling of unconventional superconductors. Therefore, these realistic self-consistent calculations capture the essence of the experiments and embrace the same spirit of the T-matrix calculation, unambiguously demonstrating the scattering nature of Co in iron-based superconductivity. Our systematic experimental-theoretical analysis of the impurity effect from a single impurity to the finite density case microscopically uncovers that the Born-limit nonmagnetic scattering is the driving force of the superconducting quantum phase transition in $LiFe_{1-x}Co_xAs$. Future characterization of the Co impurity effect by Bogoliubov quasi-particle interference imaging will be important for further exploring the orbital and band selectivity of the Born-limit nonmagnetic scattering.

**Acknowledgments:** Experimental and theoretical work at Princeton University was supported by the Gordon and Betty Moore Foundation (GBMF4547/ Hasan) and the United States Department of energy (US DOE) under the Basic Energy Sciences programme (grant number DOE/BES DE-FG-02-05ER46200). M.Z.H. acknowledges support from Lawrence Berkeley National Laboratory and the Miller Institute of Basic Research in Science at the University of California, Berkeley in the form of a Visiting Miller Professorship. We also acknowledge Korea NRF (Grant No. 2016-R1A2B4-008758), the Natural Science Foundation from Jiangsu Province of China (Grant No. BK20160094). Computations for this work were partially done with resources of Leipzig University Computing Centre. Z.W and K.J. acknowledge US DOE grant DE-FG02-99ER45747.

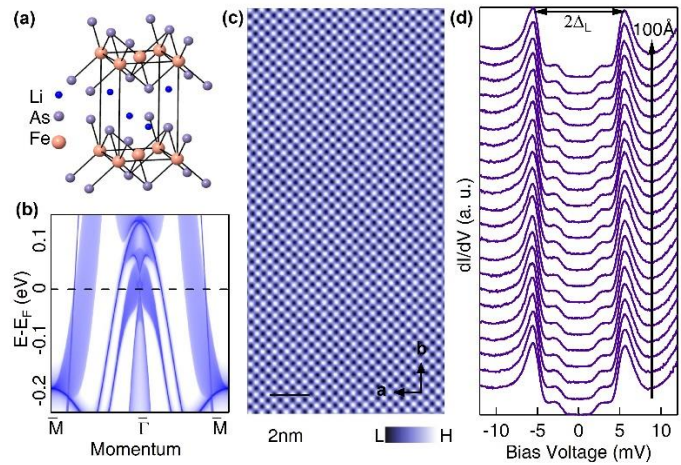

**Fig. 1.** (a) Crystal structure of LiFeAs. (b) First-principles calculation of the band structure for (001) surface. The zoom-in image shows the two Dirac cones at the zone center, with the upper one from bulk and the lower one form the surface. (c) Atomically-resolved topographic image of pristine LiFeAs showing clean tetragonal lattice. (d) Line-cut differential conductance spectra on pristine LiFeAs, showing a spatially homogeneous double-gap structure.

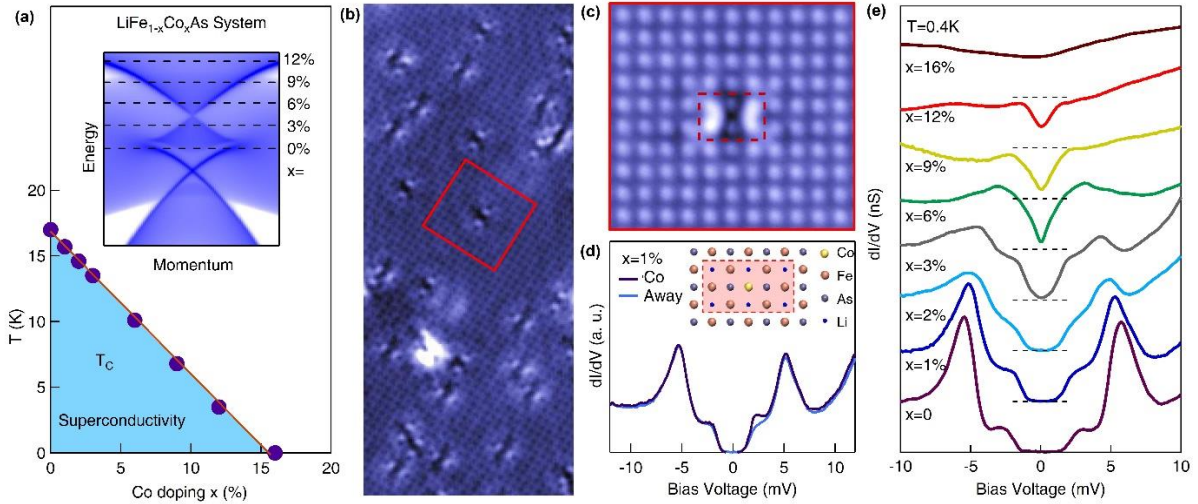

**Fig. 2.** (a) Phase diagram of LiFe$_{1-x}$Co$_x$As. The superconducting transition temperature is determined by the onset of zero resistivity. Inset: illustration of the Co doping effect on the bulk Dirac cone based on Ref. 12. (b) Atomically resolved topographic image of a sample with 1% Co substitution, showing randomly scattered dumbbell-like defects that do not exist in the pristine sample and with concentration consistent with the Co substitution level. (c) Enlarged image of single reproducible dumbbell-like defect. The center of the defect geometrically corresponds to a Co substitution atom in the Fe layer (in reference to Fig. 2D inset). (d) Differential conductance spectrum taken at the defect and far from the defect. Inset: crystal structure from top view. (e) Co concentration dependence on spatially averaged superconducting gap structure. The spectra are offset for clarity. The dashed lines mark the zero-intensity value for each case. 30 to 50 dI/dV curves taken away from apparent surface impurities with the same junction set up (V = -15mV, I= 750pA) were averaged to obtain the dI/dV curve for each concentration.

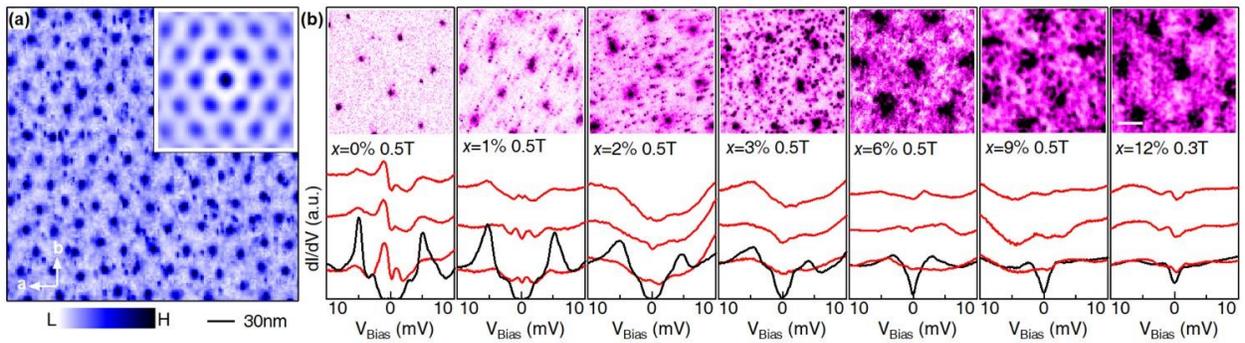

**Fig. 3.** (a) Left: real space mapping of vortices at the Fermi energy on pristine LiFeAs at $B = 2$T. Inset: auto-correlation of vortex mapping showing hexagonal lattice symmetry. (b) Spectra in the zero-field state (black) and at three representative vortices offset for clarity (red) for each concentration. The inset image in each panel shows the respective vortex lattice (the bar marks a length of 35nm).

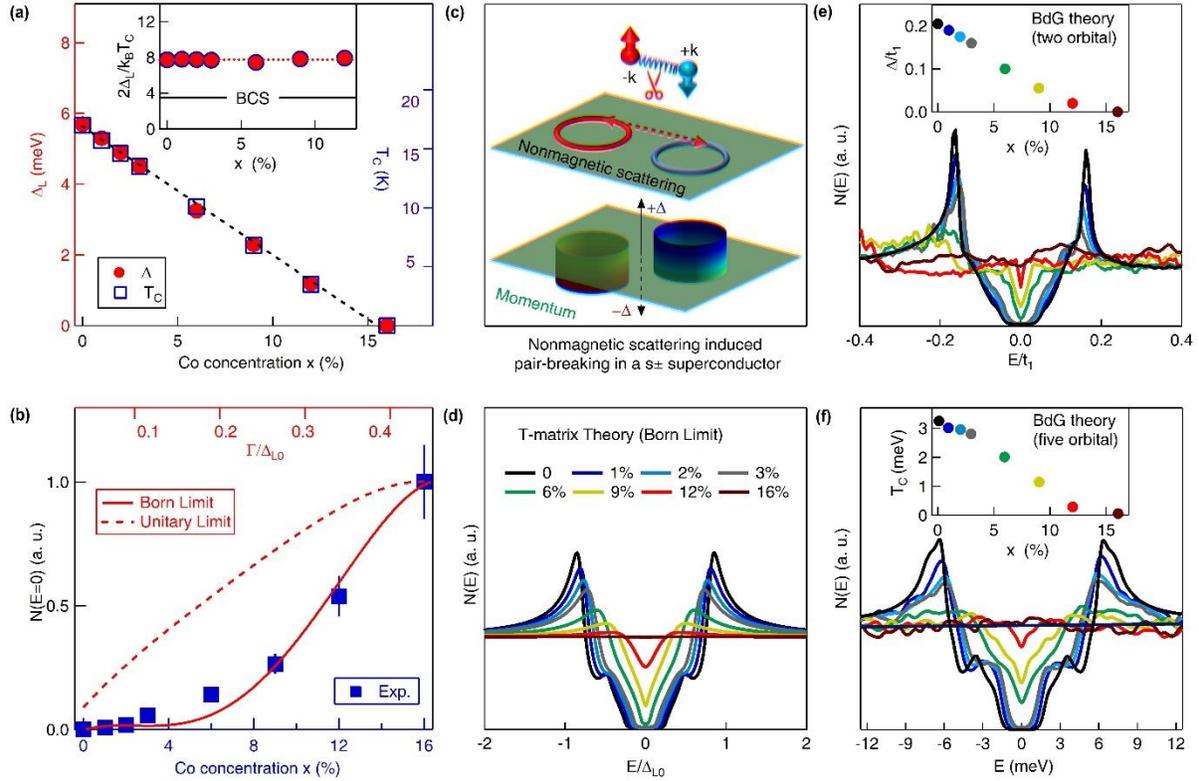

**Fig. 4.** (a) The large gap size $\Delta_L$ (left axis, red) and $T_c$ (right axis, blue) both decrease linearly as a function of concentration $x$. Inset: $2\Delta_L/k_B T_C$ remains constant (~7.7) as a function of Co concentration. (b) Differential conductance at zero energy N(E=0) as a function of Co concentration $x$ in LiFe$_{1-x}$Co$_x$As. The experimental data is normalized by the normal state value. The red solid and dashed lines denote N(E=0) calculated based on Born and unitary limit scattering, respectively. (c) Schematic showing a sign reversal s-wave pairing on two Fermi surfaces (lower panel, s± gap symmetry) and the nonmagnetic impurities induced interband scattering causing pair-breaking (upper panel). (d) Calculated density of states evolution of the s± pairing state with nonmagnetic scattering at the Born limit with T-matrix theory. (e) (f) Calculated averaged DOS evolution with increasing Co concentration by BdG theory with two-orbital and five-orbital models, respectively. The inset shows the phase diagram plot.

# Supplementary Materials

**Materials and Methods**

Single crystals of LiFe$_{1-x}$Co$_x$As grown using the self-flux method of up to 5mm × 5mm × 0.5mm were used in this study. All preparation work was carried out in an Ar filled glove box in order to protect the samples from air. Samples were cleaved mechanically *in situ* at 77K in ultra-high vacuum conditions, and then immediately inserted into the STM head, already at He4 base temperature (4.2K). The STM head that includes the sample was subsequently cooled to 0.4K with He3 cooling and stabilized, after which the magnetic field was slowly applied, with maximum temperature fluctuations of 0.2K during ramping. We waited for 1 h before performing spectroscopic imaging so that there was no noticeable vortex creep in the differential conductance map. This zero-field-cooling technique was adopted throughout this work. Tunneling conductance spectra were obtained with an Ir/Pt tip using standard lock-in amplifier techniques with a root mean square oscillation voltage of 100μV and a lock-in frequency of 973Hz. The conductance maps are taken with tunneling junction set up: V = -15mV, I = 50-150pA, while the tunneling spectra are taken with junction set up: V = -15mV, I = 750pA.

High-quality single crystals of LiFe$_{1-x}$Co$_x$As are grown with the self-flux method. The precursor of Li$_3$As is prepared by sintering Li foil and an As lump at about 700°C for 10 h in a Ti tube filled with Ar atmosphere. Fe$_{1-x}$Co$_x$As is prepared by mixing the Fe, Co, and As powders thoroughly, and then sealed in an evacuated quartz tube, and sintered at 700°C for 30 h. To ensure the homogeneity of the product, these pellets are reground and heated for a second time. The Li$_3$As, Fe$_{1-x}$Co$_x$As, and As powders are mixed according to the elemental ratio Li(Fe$_{1-x}$Co$_x$)$_{0.3}$As. The mixture is put into an alumina oxide tube and subsequently sealed in a Nb tube and placed in a quartz tube under vacuum. The sample is heated at 650°C for 10 h and then heated up to 1000°C for another 10 h. Finally, it is cooled down to 750°C at a rate of 2°C per hour. Crystals with a size up to 5 mm are obtained. The entire process of preparing the starting materials and the evaluation of the final products are carried out in a glove box purged with high-purity Ar gas. The molar ratio of Co and Fe of the LiFe$_{1-x}$Co$_x$As single crystals is checked by energy-dispersive x-ray spectroscopy (EDS) at several points on one or two selected samples for each Co concentration. For each doping, the Co concentration measured by EDS is consistent with the nominal value.

**Nonmagnetic nature of Co dopants**

**Magnetization characterization**

We measure the magnetic susceptibility of LiFe$_{1-x}$Co$_x$As using a vibrating sample magnetometer with a magnetic field of 1T to study their effective magnetic moment. As shown in Fig. S1, the measured magnetic susceptibility for different concentrations are within the same order of magnitude. Their low temperature magnetization can be described by the Curie–Weiss law [31]: $\frac{1}{\chi - \chi_0} = (T + T_\theta)/C$, and from the fitting parameter $C = \mu_0 \mu_{eff}^2 / 3k_B$ we can extract the effective local magnetic moment per Fe/Co as shown in the inset of Fig. S1. We find that in contrast to the giant enhancement of local moment for Mn and V dopants [31,25], the local moment for LiFe$_{1-}$ $_x$Co$_x$As fluctuates around 0.2$\mu_B$ per Fe/Co and the Co dopants do not substantially enhance the effective moment, thus it is more suitable to treat them as nonmagnetic impurities.

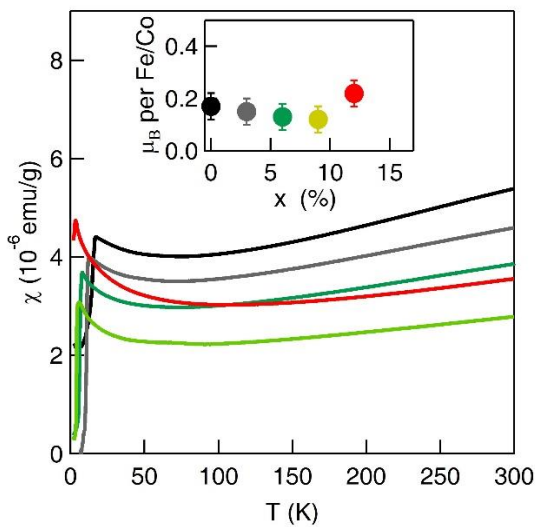

**Fig. S1** Temperature dependence of the dc magnetic susceptibility in a 1 T magnetic field for different Co concentrations. The inset shows the extracted effective local magnetic moment.

**First-principles calculation**

First-principles calculations were performed in the density functional theory [32,33] framework as implemented in the Vienna Ab initio Simulation Package (VASP) [34]. Generalized gradient approximation in Perdew−Burke−Ernzerhof (PBE) functional [35][36] was applied to describe electron exchange-correlation interaction with the projector augmented wave (PAW) potentials [37]. The energy cutoff was set at 500 eV. The energies in self-consistent calculations were converged until $10^{-5}$ eV. Striped antiferromagnetic, ferromagnetic and non-magnetic states are simulated via non-collinear self-consistent calculations of $\sqrt{2} \times \sqrt{2} \times 1$ LiFeAs supercell for undoped and Co-doped, LiFe$_{(1-x)}$Co$_x$As where x = 0.5. The Brillouin zone was sampled using a 16 x 16 x 10 Monkhorst-Pack [38] grid.

We explored the effect of Co substitution on the magnetism of LiFeAs, and investigated ferromagnetic (FM), striped antiferromagnetic (AFM), and nonmagnetic (NM) orientations. DFT calculations show that Co substitution suppresses the magnetism in LiFeAs. In Table S1, undoped LiFeAs have three distinct magnetic configurations – FM, AFM and NM. Striped AFM is the most stable magnetic orientation, with 0.156 eV and 0.167 eV per $\sqrt{2}$ supercell lower than FM and NM cases. The non-striped type of antiferromagnetic configuration is degenerate with the NM case. The system energies and magnetic moments are summarized in Table S1.

However, after partial Co substitution, the system becomes unstable as the magnetic states are no longer distinguishable. From Table S2, we can find that the ferromagnetic states are suppressed, and the striped AFM states become degenerate with NM states. At concentration x = 0.5, two possible arrangements of Co doping arise, i.e., (i). Fe atoms with two and (ii). four nearest-neighbor Co atoms denoted as A1(Linear) and A2 (Alternating), respectively. The A2 configuration is generally the preferred atomic arrangement by 0.05 eV per $\sqrt{2}$ supercell. In these configurations, the FM calculations converge to NM and the net magnetic moments become zero.

The AFM orientations shown in Fig. S2 present two types of striped AFM, labeled as parallel and crossed. Parallel stripes occur when the magnetic moments are along a line of nearest-neighbor doped atoms having the same direction. On the other hand, in the crossed stripes orientation, the nearest-neighbor doped atoms have an opposite direction from each other. This is schematically shown in the rightmost panel of Fig. S2. Table S2 shows that the energies of both types of striped AFM and NM cases are degenerate.

Examining the local magnetic moments, we found that Co doping reduced the magnetic moments of each atom as listed in Table S3. Specifically, in the linear Co doping arrangement, A1, with crossed stripes, the local magnetic moment of Co and Fe atoms are $0.071\mu_B$ and $0.381\mu_B$, respectively. This is a substantial reduction as compared to the undoped LiFeAs where Fe atoms have a larger local magnetic moment of $1.213\mu_B$. Clearly, these results confirm that in partially Co-doped LiFeAs, Co dopants exhibit a non-magnetic nature.

Finally, we estimate the on-site potential of Co dopants. We first calculate the orbital resolved DOS for LiFe$_{0.5}$Co$_{0.5}$As. As shown in Fig. S3, the DOS of Co $3d$ orbitals overlaps substantially with that of Fe $3d$ orbitals without apparent sharp bound state, indicating weak potential scattering nature of the Co dopants. We then calculate the "center of mass" for each partial DOS. The energy for the center of mass is used to estimate the on-site energy of the orbital as $U=\int_{-6eV}^{+3eV} DOS(E)E / \int_{-6eV}^{+3eV} DOS(E)$, where DOS(E) is the partial density of state as a function of E. The calculated on-site energies for Co $3d$ and Fe $3d$ as -1.52eV and -1.09eV, respectively. The on-site potential of Co is then estimated by their difference to be -0.43eV, which is of the same sign and order of magnitude with previous on-site potential estimation [39] of Co dopants in LaOFeAs.

| Magnetic Orientation | Undoped LiFeAs | |
|---|---|---|
| | Energy (eV) | Mag. ($\mu_B$) |
| **Striped AFM** | -64.816 | 0.000 |
| **FM** | -64.660 | 1.338 |
| **NM** | -64.649 | 0.000 |

**Table S1.** System energies and net magnetic moments undoped LiFeAs with different magnetic orientations.

| | Arrangement of Co and Fe atoms | | | |
| --- | --- | --- | --- | --- |
| **Magnetic Orientation** | **A1 - Linear** | | **A2 - Alternating** | |
| | **Energy (eV)** | **Mag. (μB)** | **Energy (eV)** | **Mag. (μB)** |
| **FM** | *converged to NM* | | *converged to NM* | |
| **AFM Stripes** **Parallel** | -62.812 | 0.000 | -62.863 | 0.000 |
| **AFM Stripes** **Crossed** | -62.811 | 0.000 | -62.863 | 0.026 |
| **NM** | -62.811 | 0.000 | -62.863 | 0.000 |

**Table S2.** System energies and net magnetic moments for partial Co substitution, x = 0.5, with different magnetic orientations.

| | **Antiferromagnetic orientation** | **Magnetic moment per atom (μB)** | |
| --- | --- | --- | --- |
| **Co Concentration, x** | | **Fe** | **Co** |
| **0.0** | Undoped LiFeAs | 1.213 | N/A |
| **0.5** | Linear – parallel stripes | 0.003 | -0.003 |
| **0.5** | Linear – crossed stripes | ±0.381 | ±0.071 |
| **0.5** | Alternating – parallel stripes | ±0.043 | 0.001 |
| **0.5** | Alternating – crossed stripes | 0.035 | -0.022 |

**Table S3.** Local magnetic moments of Fe and Co atoms for undoped and Co-doped LiFeAs.

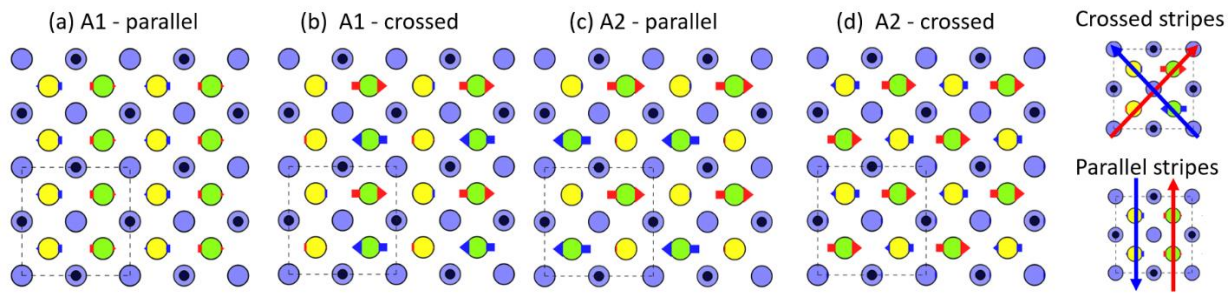

**Figure S2.** The non-collinear striped AFM orientations of LiFeAs with Co concentration of x = 0.5. The parallel and crossed AFM stripes of linearly (A1) arranged Co and Fe atoms are shown in (a) and (b), respectively. The 3$^{rd}$ and 4$^{th}$ panels show the alternating (A2) arrangement for (c)

parallel and (d) crossed AFM strips. The rightmost panel shows the schematic difference between crossed and parallel AFM stripes.

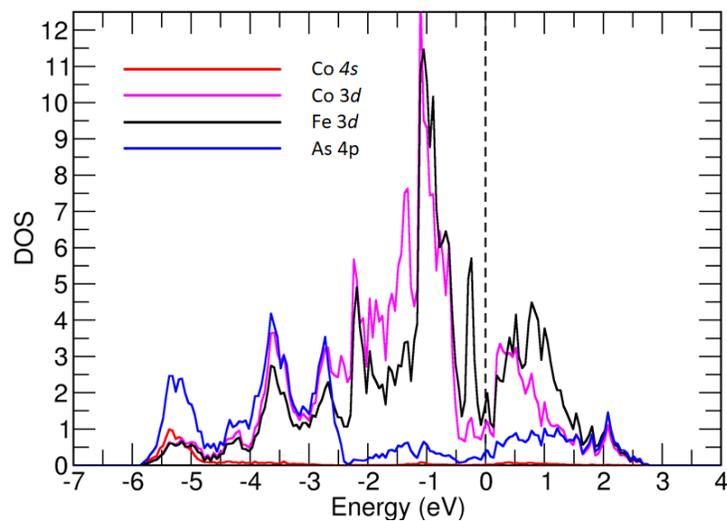

**Figure S3.** Calculated orbital resolved DOS for LiFe$_{0.5}$Co$_{0.5}$As.

**Theoretical simulation finite density of Co on the superconducting ground state**
**T-matrix calculation**

We use the $\mathcal{T}$-matrix approximation [26,40] generalized to the $s^{\pm}$-wave state of the two band model, to study the effects of impurities on the evolution of the experimentally measured local DOS with a systematic Co doping. In this study we freely tuned the strength of the impurity potential from the weak (Born limit) to the strong (unitary limit) coupling to fit the experimental data. We also tested a magnetic impurity potential on the assumed $s^{++}$-wave state, and found that the weak non-magnetic impurity scatterers on the $s^{\pm}$-wave state can describe the experimental DOS $N(\omega)$ and its evolution with Co-doping.

The impurity induced self-energies are calculated with the $\mathcal{T}$-matrix generalized to a two-band superconductivity as

$$T_a^i(\omega_n) = \frac{G_a^i(\omega_n)}{D} \quad (i=0,1; a=h,e) \quad (1)$$

$$D = c^2 + |G_h^0 + G_e^0|^2 + |G_h^1 + G_e^1|^2 \quad (2)$$

$$G_a^0(\omega_n) = \frac{N_a}{N_{tot}} \left\langle \frac{\widetilde{\omega_n}}{\sqrt{\widetilde{\omega_n^2} + \widetilde{\Delta_a^2}(k)}} \right\rangle \quad (3)$$

$$G_a^1(\omega_n) = \frac{N_a}{N_{tot}} \left\langle \frac{\widetilde{\Delta_n}}{\sqrt{\widetilde{\omega_n^2} + \widetilde{\Delta_a^2}(k)}} \right\rangle \quad (4)$$

where $\omega_n = T\pi(2n+1)$ is the Matsubara frequency, and $N_{tot} = N_h(0) + N_e(0)$ is the total DOS. $c = \cot\delta_0 = \frac{1}{\pi N_{tot} I_{imp}}$ is a convenient measure of scattering strength $I_{imp}$, with $c = 0$ for the unitary limit and $c > 1$ for the Born limit scattering. $\langle ... \rangle$ denotes the Fermi surface average. The subscript $a$ stands for the electron band and hole band, respectively, and the superscript $i$ stands for the normal ($i = 0$) and anomalous ($i = 1$) part of Green's functions, respectively.

The above four $\mathcal{T}$-matrices, $\mathcal{T}_a^i$, are numerically solved and the corresponding impurity induced self-energies are obtained as

$$\Sigma_{h,e}^{0,1}(\omega_n) = \Gamma \cdot T_{h,e}^{0,1}(\omega_n), \quad \Gamma = \frac{n_{imp}}{\pi N_{tot}} \quad (5)$$

where $\Gamma$ is the impurity concentration parameter with $n_{imp}$ the impurity density per unit cell. The normal/amomalous self-energy corrections to the Green's functions are then:

$$\widetilde{\omega_n} = \omega_n + \Sigma_h^0(\omega_n) + \Sigma_e^0(\omega_n) \quad (6)$$

$$\widetilde{\Delta_{h,e}} = \Delta_{h,e} + \Sigma_h^1(\omega_n) + \Sigma_e^1(\omega_n) \quad (7)$$

.
The important part for an impurity bound state is $D$ in Eq.(2). Being a denominator of $\mathcal{T}$-matrices, $\mathcal{T}_a^i$, it signals a formation of a bound state when it goes to zero; otherwise, no bound state exists. The last term in $D$, namely, $|G_h^1 + G_e^1|$ would exactly vanish for a d-wave superconductor because the FS average over the d-wave order parameter becomes zero, hence a zero energy bound state forms when $c = 0$. For the s±-wave case, a cancellation still occurs because $G_h^1$ and $G_e^1$ have opposite signs. However, this cancellation is never perfect unless $|\Delta_e| = |\Delta_h|$ and $N_h(0) = N_e(0)$. With an incomplete cancellation, this finite remnant $|G_h^1 + G_e^1|$ acts as a weakening impurity scattering strength (increasing the effective value of $c$). Therefore, the impurity bound state in the s±-wave state forms at finite energies symmetrically split relative to the zero energy even with unitary impurity $c = 0$. Decreasing the impurity scattering strength towards a Born limit, these split in-gap bound states move towards the gap edges and merge to the quasi-particle continuum.

By fitting the experimental DOS (we use $N_L=N_e$, and $N_S=N_h$ as in the main text), we can effectively determine the nature of the impurities and its coupling strength. We set realistic parameters: DOS ratio: $N_L/N_S=2.5$; gap size ratio: $\Delta_S/\Delta_L=-0.55$, with a sign reversal; the scattering rate is set to be proportional to the Co concentration with a small offset: $\Gamma/\Delta_{L0}=0.05+0.4(x/x_C)$ where $x_C=16\%$; and $\Delta_L=\Delta_{L0}(x_C-x)/x_C$ in reference to Fig. **4a**. Figure S4 shows the quantum many body self-energy (the imaginary part) for different Co concentrations at the Born limit.

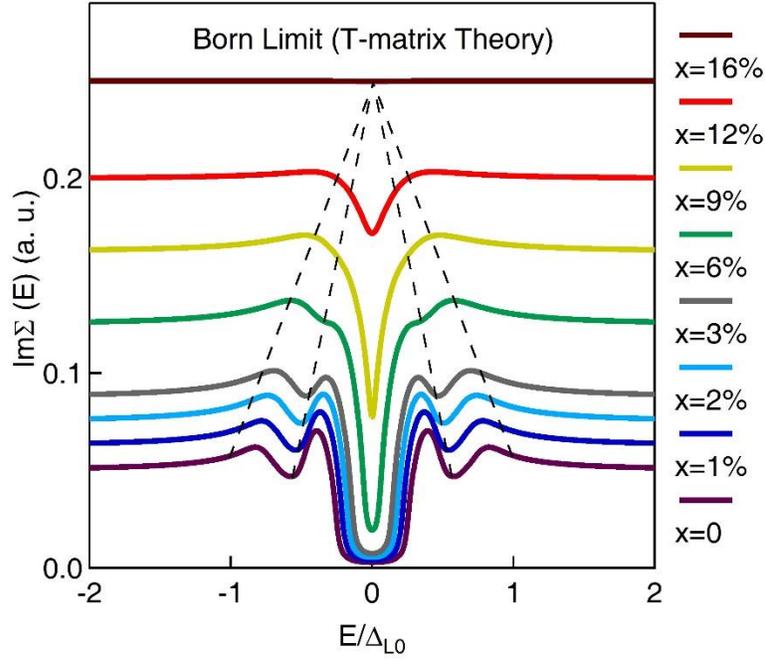

**Figure. S4** Nonmagnetic impurity induced many-body self-energies (imaginary part) for different concentrations at the Born limit. The dashed lines illustrate the reduction of the two gaps.

## Bogoliubov–de Gennes self-consistent calculation

### Two-orbital model

We adopted a two-orbital tight-binding model proposed in Ref. 28. Based on this model, there are many numerical results [41-44] consistent with experiments. Thus, this model may be a starting point to study the low-energy excitations for the iron-based superconductors. Here we also apply this model while considering the results from the photoemission experiment and the first-principles calculations in Co-doped LiFeAs.

This model is given by

$$H = -\sum_{i\mu j\nu\sigma} \left(t_{i\mu j\nu} c_{i\mu\sigma}^\dagger c_{j\nu\sigma} + H.c.\right) - \mu_c \sum_{i\mu\sigma} c_{i\mu\sigma}^\dagger c_{i\mu\sigma}$$
$$+ \sum_{i\mu j\nu\sigma} \left(\Delta_{i\mu j\nu} c_{i\mu\sigma}^\dagger c_{j\nu\bar\sigma}^\dagger + H.c.\right) + U \sum_{i\mu\sigma\neq\bar\sigma} \langle n_{i\mu\bar\sigma}\rangle n_{i\mu\sigma} + U' \sum_{i,\mu\neq\nu,\sigma\neq\bar\sigma} \langle n_{i\mu\bar\sigma}\rangle n_{i\nu\sigma}$$
$$+ (U' - J_H) \sum_{i,\mu\neq\nu,\sigma} \langle n_{i\mu\sigma}\rangle n_{i\nu\sigma} + \sum_{i_m\mu\sigma} V_{imp}\, c_{i_m\mu\sigma}^\dagger c_{i_m\mu\sigma}$$

(8)

Where $i = (i_x, i_y), j = (j_x, j_y)$ are the site indices in two-dimensional plane, $\mu, \nu = 1, 2$ are the orbital indices, and $n_{i\mu\sigma}$ is the density operator at site $i$ and orbital $\mu$, $U$ ($U'$) is the on-site intraorbital (interorbital) Coulomb interaction and $J_H$ is the Hund's rule coupling. The quantity $U'$ is taken to be $U - 2J_H$, assuming the orbital rotation symmetry [43]. For a nonmagnetic impurity located at site $i_m$, we consider the intra-orbital scattering with the strength $V_{imp}$. In addition, $\mu_c$ is the chemical potential, which is determined by the electron filling, corresponding to different doping level $x$. Considering the two-orbital tight-binding model here, $x$ is related to the band filling as $n = 2 + x$. The hopping constants $t_{i\mu j\nu}$ are chosen as follows:

$$t_{i1,i\pm\hat{x}1} = t_{i2,i\pm\hat{y}2} = t_1,$$

$$t_{i1,i\pm\hat{y}1} = t_{i2,i\pm\hat{x}2} = t_2,$$

$$t_{i\mu,i\pm(\hat{x}+\hat{y})\mu} = \frac{1+(-1)^i}{2}t_3 + \frac{1-(-1)^i}{2}t_4, \quad (9)$$

$$t_{i\mu,i\pm(\hat{x}-\hat{y})\mu} = \frac{1+(-1)^i}{2}t_4 + \frac{1-(-1)^i}{2}t_3,$$

$$t_{i\mu,i\pm\hat{x}\pm\hat{y}\nu} = t_5, (\mu \neq \nu)$$

The mean-field Hamiltonian can be diagonalized by solving the Bogoliubov-de-Gennes (BdG) equations,

$$H = \sum_j \sum_\nu \begin{pmatrix} H_{i\mu j\nu\sigma} & \Delta_{i\mu j\nu} \\ \Delta^*_{i\mu j\nu} & -H^*_{i\mu j\nu\bar{\sigma}} \end{pmatrix} \begin{pmatrix} u^n_{j\nu\sigma} \\ v^n_{j\nu\bar{\sigma}} \end{pmatrix} = E_n \begin{pmatrix} u^n_{i\mu\sigma} \\ v^n_{i\mu\bar{\sigma}} \end{pmatrix} \quad (10)$$

where

$$H_{i\mu j\nu\sigma} = -t_{i\mu j\nu} + \left(U\langle n_{i\mu\bar{\sigma}}\rangle + (U-2J_H)\langle n_{i\bar{\mu}\bar{\sigma}}\rangle + (U-3J_H)\langle n_{i\bar{\mu}\sigma}\rangle + V_{imp}\delta_{i,i_m} - \mu_c\right)\delta_{ij}\delta_{\mu\nu} \quad (11)$$

and

$$\Delta_{i\mu j\nu} = \frac{V_{i\mu j\nu}}{4}\sum_n (u^n_{i\mu\uparrow}v^{n*}_{j\nu\downarrow} + u^n_{j\nu\uparrow}v^{n*}_{i\mu\downarrow}) \tanh\left(\frac{E_n}{2k_BT}\right) \quad (12)$$

$$\langle n_{i\mu\uparrow}\rangle = \sum_n |u^n_{i\mu\uparrow}|^2 f(E_n) \quad (13)$$

$$\langle n_{i\mu\downarrow}\rangle = \sum_n |v^n_{i\mu\downarrow}|^2 (1-f(E_n)) \quad (14)$$

$$\langle n_{i\mu}\rangle = \langle n_{i\mu\uparrow}\rangle + \langle n_{i\mu\downarrow}\rangle \quad (15)$$

here, $f(E_n)$ is the Fermi-Dirac distribution function, and $V_{i\mu j\nu}$ is the pairing strength. Here we consider the $s_{\pm}$- wave symmetry (see main text) and choose the next-nearest-neighbor (NNN) intraorbital pairing with strength $V_{i\mu j\nu} = V_{ij} = V_{NNN}$ as a constant [21-26, 44]. In addition, we define the local magnetization and $s_{\pm}$- wave projection of the superconductivity order parameter at each site $i$, respectively as: $m_i = \frac{1}{2}\sum_\mu(\langle n_{i\mu\uparrow}\rangle - \langle n_{i\mu\downarrow}\rangle)$, $\Delta_i = \frac{1}{8}\sum_{\mu\delta}\Delta_{i\mu,i+\delta\mu}$, where $\delta = \pm\hat{x} \pm \hat{y}$. When determining the strength of the pairing symmetry for a given different doping level $x$, we take an average over the whole lattice positions and disorder configurations for each local pairing amplitude shown in Eq. (12).

Throughout this work, the energies are measured in units of $t_1$, the temperature is set to be $T = 0.0001$, the hopping constants are chosen as $t_{1-5} = (1, 0.7, 0.5, -2.0, 0.16)$. The band energy and fermi surface without interaction has been depicted in Fig. S5. With electron doping the Fermi surface nesting condition is enhanced, consistent with the photoemission data in Ref. 9. The intraorbital Coulomb interaction $U$ and the pairing strength $V_{NNN}$ are set to be 3.4 and 1.4, respectively, the Hund's rule coupling $J_H = U/4$. Based on our first-principles calculation mentioned above, the on-site potential of Co is estimated to be as weak as -0.43eV. In model calculation, a further renormalization factor around 2 is often used to taking the correlations into account [19]. Thus here $V_{imp}$ is set to be -2 (which amounts to ~ -0.2eV much smaller than the total bandwidth ~1.2eV in the model). With these realistic parameters, we calculated the BdG equations self-consistently with different doping level. The numerical calculations are performed on a $28 \times 28$ square lattice with periodic boundary conditions. At each doping level, the calculations are performed on 25 different configurations, in each of which Co dopants are distributed randomly and homogeneously. Co dopants not only provide the onsite scattering, but also contribute extra electrons into the system. With these considerations, we obtain the linear decreasing trend of the superconductivity order parameter with increasing Co concentration, as shown in Fig. 4. To investigate the local dopant effect, we calculate the local DOS at $x = 1\%$ around the Co dopant, and compare the results with that far away from the dopant (Fig. S6). The LDOS can be expressed as

$$\rho_i(\omega) = \sum_{n,\mu}\left[\left|u_{i\mu\sigma}^n\right|^2\delta(E_n - \omega) + \left|v_{i\mu\bar\sigma}^n\right|^2\delta(E_n + \omega)\right] \quad (16)$$

where the $\delta$ function is taken as $\Gamma/\pi(x^2 + \Gamma^2)$, with the quasiparticle damping $\Gamma = 0.003$. In addition, the averaged DOS at each doping level are calculated. A $32 \times 32$ supercell is used to calculate the averaged DOS.

In the presence of a magnetic field $B$ perpendicular to the plane, the hopping integral can be expressed as $t'_{i\mu j\nu} = t_{i\mu j\nu}\exp\left[i(\pi/\Phi_0)\int_i^j \vec{A}(\vec{r})\cdot d\vec{r}\right]$, where $\Phi_0 = h_c/2e$ is the superconducting flux quantum, and $\vec{A}(\vec{r}) = (-By, 0, 0)$ is the vector potential in the Landau gauge. In our calculation, magnetic unit cells are introduced where each unit cell accommodates two superconducting flux quantum and the linear dimension is $N_x \times N_y = 64 \times 32$. A $16 \times 32$ supercell is used to calculate the local density of states. The vortex core state is shown in Fig. S6 inset.

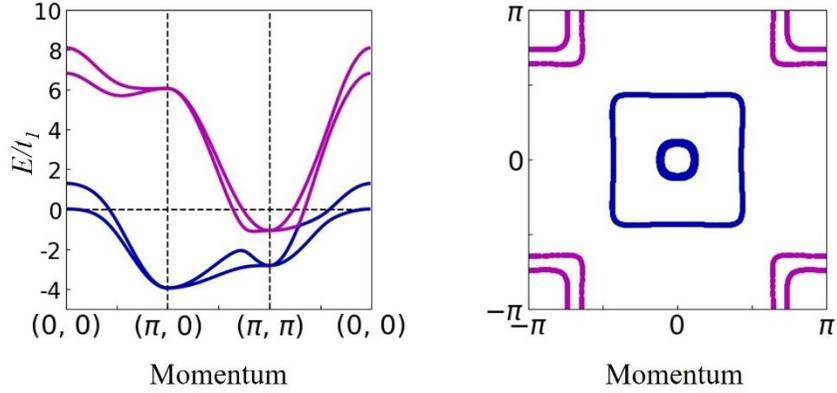

**Figure S5.** The two-orbital model calculated band structure (left) and Fermi surface (right) for LiFeAs.

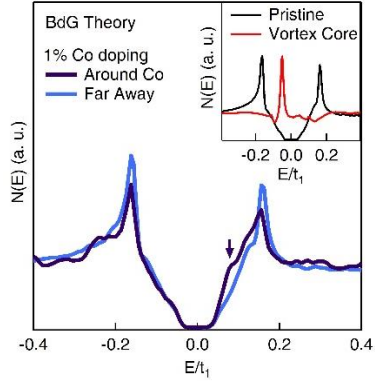

**Figure S6.** The two-orbital model calculated single Co impurity effect and vortex core states.

## Five-orbital model

We use the tight binding model as deduced earlier from spectral positions of the band structure as measured in photoemission [45]. This model was obtained by fitting the symmetry allowed hoppings [46] at short ranges such that the orbital content at the Fermi level matches experimental evidences as well. The band structure and the Fermi surface of that model for the pristine LiFeAs is presented in Fig. S7. The superconducting order parameter has been obtained self-consistently using a real space implementation of the BdG approach using pairing interactions in real space (where the pairing has been cut at a distance of three lattice spacings in x and y direction) that have been calculated from a modified spin-fluctuation approach [47] within the same tight binding model. Upon Co doping, the pairing interaction itself is kept constant. For the homogeneous case one obtains a superconducting order parameter with a structure as shown in Fig. S8 (where the real space structure, its Fourier transform and the corresponding projection to the Fermi surface is presented). Next, random impurity configurations are taken to simulate Co substituting for Fe in the system. The impurity potential is chosen to be V imp = −0.15eV, which is in agreement with the value as found from ab initio calculations by taking into account a quasiparticle renormalization factor of Z=1/2 [19].

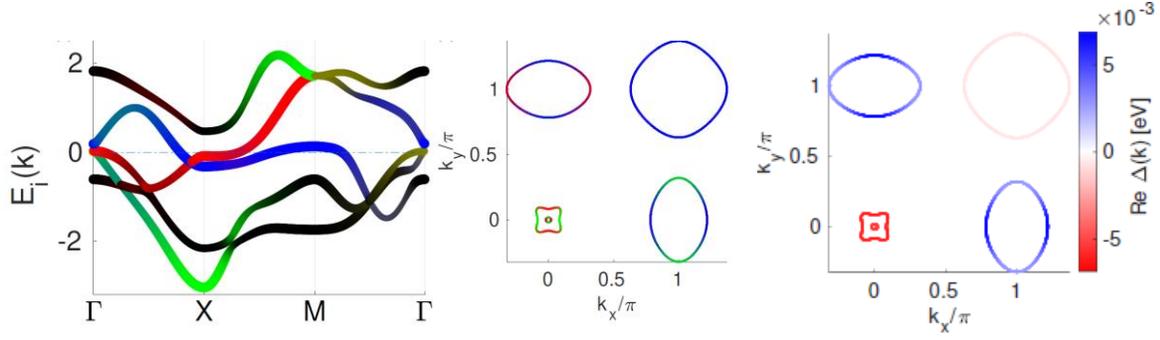

**Figure S7.** Bands along high symmetry directions together with the orbital character (left) and Fermi surface of the 2D version of the electronic structure (middle) and gap structure (right) for LiFeAs. Color code: red dxz, green dyz, blue dxy, black (other).

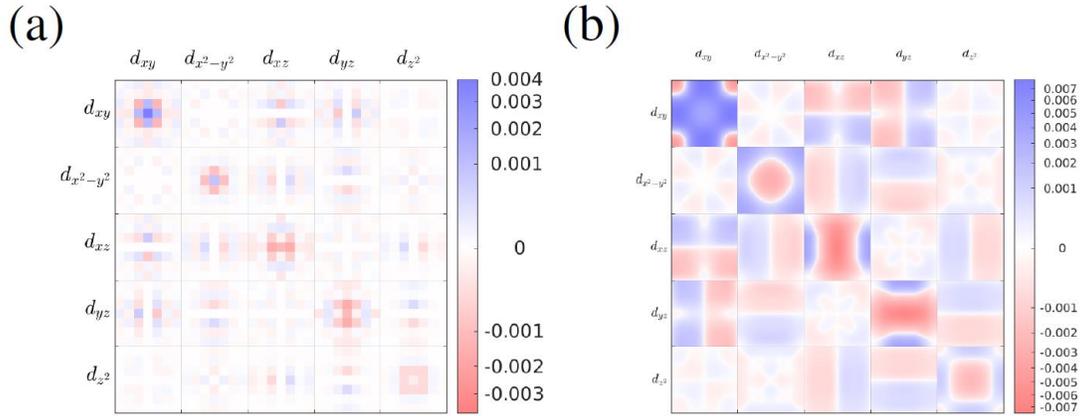

**Figure S8.** Plot of the mean fields as obtained in the self-consistent calculation of a homogeneous system. Structure of gap for all combinations of the orbitals in the real space (a) and in momentum space (b). Projection of the order parameter in band space shown on the Fermi surface.

**Further discussion of pairing on the bulk Dirac bands**

In addition, the electron doping also causes the system's Fermi level to cross the bulk Dirac cone (Fig. 2**a** inset) and there will be two corresponding spherical Fermi surfaces along the Γ-Z direction [12]. Due to the intrinsic orbit-momentum locking in such Fermi surfaces [48,49], its gap function can be either nodeless or nodal. When we assume its pairing is induced from the s± state of the ordinary bands, the most natural pairing on the bulk Dirac bands should be spin-singlet and intra-orbital, which is a s-wave gap. In principle, the spin triplet inter-orbital pairing is also allowed for the bulk Dirac Fermi surfaces, and the associated gap function has point nodes along the $k_z$ axis [48,49], which is incompatible with the s± gap function. The frustration in pairing symmetry, in this case, can suppress the Cooper pairing, and its impact can be non-monotonic when the Fermi level systematically crosses the bulk Dirac cone via Co doping. However, experimentally both the gap and $T_C$ are linearly suppressed, which does not directly support the latter case. Therefore, we conclude that the main source of the linear $T_C$ reduction may still come from a finite density of nonmagnetic scatters in a s± superconductor.